\documentclass[preprint]{revtex4} % PAPIERFORMAT, STANDARDSCHRIFTGRÖSSE, DOKUMENTENART

\usepackage{eurosym} % ZUR VERWENDUNG DES EURO- SYMBOLS
\usepackage{amsmath} % FÜR MATHEMATISCHE SYMBOLIK
\usepackage{amssymb} % FÜR MATHEMATISCHE SYMBOLIK
\usepackage{dsfont} % FÜR MATHEMATISCHE SYMBOLIK
\usepackage{enumerate} % FÜR AUFZÄHLUNGEN

\usepackage{graphicx} % ZUM GRAPHIKEN EINBINDEN

\usepackage[mathscr]{eucal} % FÜR MATHEMATISCHE SYMBOLIK
\usepackage{longtable} % ZUM ERSTELLEN VON SEHR LANGEN TABELLEN
\usepackage{multirow}

\usepackage{color}

\newcommand{\dd}{\mathrm{d}}
\newcommand{\Dp}{\partial}
\newcommand{\un}{\infty}
\newcommand{\ve}{\varepsilon}

\newcommand{\li}{\left}
\newcommand{\ri}{\right}

\newcommand{\abs}[1]{\li| #1 \ri|}

\newcommand{\cen}[1]{\begin{center} #1 \end{center}}

\newcommand{\tc}{T_{{\rm c}}}

\usepackage{lscape}
\usepackage{rotating}

\begin{document}
\title{Extended soft wall model with background related to \\ features of QCD thermodynamics}

\author{R. Z\"ollner}
\author{B. K\"ampfer}
\affiliation{Helmholtz-Zentrum Dresden-Rossendorf, PF 510119, D-01314 Dresden, Germany \\ and \\ Institut f\"ur Theoretische Physik, TU Dresden, D-01062 Dresden, Germany}

\date{\today}

\begin{abstract}
The soft wall model is extended to accommodate at the same time (i) approximately linear $\rho$ meson Regge trajectories at zero temperature $T$, (ii) various options for the thermodynamics with reference to QCD (cross over or second-order transition or first-order transition at $\tc$), and (iii) the appearance of vector meson states at $T \lesssim \tc$. While the vector meson masses display some modest model dependence very near to $\tc$, they stay below $\tc$ to good accuracy independent of the temperature, that is nearly as at $T=0$, thus being very consistent with the thermo-statistical models widely employed in analyses of the hadron yields in relativistic heavy-ion collisions in a region where baryon densitiy effects can be neglected and the vacuum hadron masses are used.
\end{abstract}


\maketitle

\section{Introduction}
The soft wall (SW) model \cite{KKSS} is a particular realization of ideas rooting in the AdS/CFT correspondence \cite{Maldacena,Witten,Gubser}. It can be considered as holographic bottom-up model allowing for the access to the excitation spectrum of vector mesons. (For extensions to other meson species, see \cite{Col12,Col09, Gherghetta,BK}.) Vector mesons ($V$), such as $\rho$, $\omega$, $\phi$ and $J/\psi$, are of special interest for dilepton $(l \bar l)$ spectroscopy in relativistic heavy-ion collisions \cite{Rapp1,Rapp2} since they couple as $J^{PC} = 1^{- -}$ states to photons $V \to \gamma^{\ast} \to l \bar l$ and allow to seek for medium modifications via their spectral functions (cf. \cite{Leu, Hay, Fri, Fuk} for surveys and \cite{Hilger,Hohler} for investigations related to $\rho$ mesons). For that, the SW model has to be extended to accommodate non-zero temperature effects. This has been accomplished, e.g. in \cite{Col12}, with the finding of the disappearance of vector mesons at a temperature scale significantly below $\tc$ (cf.~also \cite{Bartz-Jacobson}), where $\tc$ refers either to the cross over temperature of QCD or the critical temperature of first-order/second-order phase transition of QCD in the chiral/nearby chiral limit. As shown in \cite{ich1}, a suitable modification of the SW model can serve for both, an accurate description of the Regge trajectories of radial excitations and a disappearance temperature of $\mathcal{O}(\tc)$. Even an extension to finite baryon density effects is possible \cite{ich2}. The construction in \cite{ich1, ich2} is tightly related to the Hawking-Page phase transition, implying a first-order phase transition at $\tc$, where at temperatures $T$ below $\tc$ the thermal gas solution applies, i.e.~the hadron states are as at $T=0$. \\
It is the aim of the present paper to report on the thermodynamics which is encoded in our extension of the SW model in relation with QCD thermodynamics and the option of tuning the holographic model such to have vector meson states only at $T \lesssim \tc$. The picture we have in mind is as follows. At $T >\tc$, no hadron states exist. At $\tc$ (or in a narrow corridor centered at $\tc$), hadron states appear and persist toward smaller temperatures. In other words, considering an expanding and therefore cooling piece of strongly interacting matter in the course of a relativistic heavy-ion collision, hadronization happens at $\tc$. Guided by the success of the thermo-statistical model \cite{Stachel2, Stachel3, CKWX, Bec2, thst1, thst3}, we argue that the emerging hadron states are statistically distributed according to maximum entropy. In line with arguments in \cite{Stachel}, the inelastic hadron reaction rates drop extremely rapidly upon further cooling, i.e.~the chemical composition freezes out (f.o.) at $T_{\rm f.o.} \approx \tc$. (For a more differential analysis within transport models, cf. \cite{Becattini}.) \\
Altogether, we provide a holographic model in the spirit of the SW model with $T_{\rm dis} \approx T_{\rm f.o.} \approx \tc$, where $T_{\rm dis}$ is the temperature at which hadron states appear upon cooling (or disappear when heating a piece of strongly interacting matter). At the same time, an approximately linear Regge trajectory with proper $\rho$ ground state at $T=0$ is used as input for the QCD-hadron scale setting. For the thermodynamic scale setting we employ $\tc \approx 150$ MeV, as turns out from lattice QCD calculations \cite{Borsanyi2014, Bazavov2014} for 2+1 flavors with physical quark masses, where the label `c' refers to `cross-over', albeit we discuss also the options of a second-order phase transition and a first-order phase transition which are enabled in 2+1 flavor QCD for special quark mass parameters according to the Columbia plot, cf.~\cite{Philipsen:2015eya}. In our model, these various cases are steered by the continuous change of a single parameter. Finally, we show that parameter tuning allows for the scale setting such to accomplish $T_{\rm dis} \cong \tc$. As a result of these scale settings we find only tiny medium modifications at $T<\tc$, here caused by the ambient hot medium wherein the selected hadrons (vector mesons) are immersed. This is in agreement with the thermo-statistical analyses of data on hadron abundances which employ successfully the hadron spectrum at $T=0$. \\
Our paper is organized as follows. In Sec.~II we introduce the extended SW model and present the vector meson spectrum at $T=0$. The thermodynamics is discussed in Sec.~III, and Section IV contains the features of the vector meson spectrum at non-zero temperature, in particular the systematics of $T_{\rm dis}$. We summarize in Sec.~V. The Appendix provides a brief recourse to the Hawking-Page transition.

\section{The extended soft wall model and setting the hadron scale}
\begin{figure}
   \cen{\includegraphics[scale=.75]{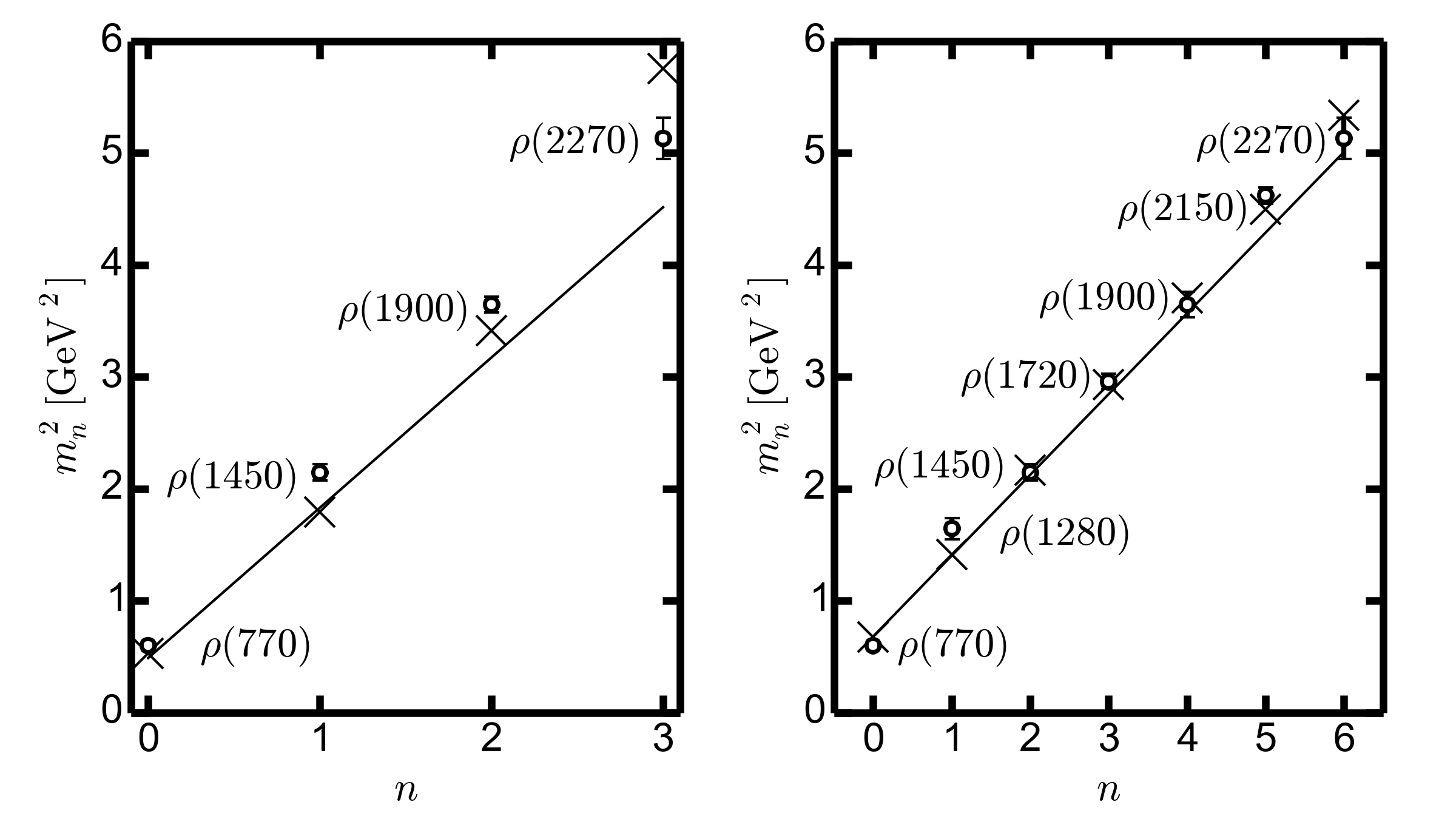}
   \caption{Selection of states (circles with error bars) belonging to the $\rho$ Regge trajectory according to \cite{Ebert} (left panel) or to \cite{BK} (right panel). The crosses exhibit solutions of Eq.~(\ref{schroe}) for $f=1$ with $c=291$ MeV, $p=3.77$ (left panel) and $c=399$ MeV, $p=2.09$ (right panel). The straight lines are for $m_n^2= \beta_0+\beta_1n $ with the values of $\beta_{0,1}$ quoted in the text.}\label{p3_abb1}}
  \end{figure}

The holographic approach to vector mesons arises from a stack of $N_c$ (number of colors) coincident $D3$-branes resulting in a ten-dimensional space-time. In the spirit of the AdS/CFT duality it is AdS$_5 \times S^5$, where AdS$_5$ is a five dimensional anti-de Sitter `universe' and $S_5$ is a compact sphere which drops out in the strong coupling limit. $N_f$ (number of flavors) coinciding $D7$-branes are then included to supply flavor gauge fields \cite{Erdemenger, Erd2, Erd3, Erd4}. This approach leads to a general action including the massless real-valued scalar dilaton background field $\Phi$, a bulk tachyon field as well as the gauge field strength tensors squared \cite{Cui11}. In the probe limit and focusing on the vector-like gauge field $V_M$ the action simplifies to \cite{KKSS}
  \begin{equation}
   S_V = \frac1k  \int  \! \dd z\, \dd^4 x \,   \sqrt{g} e^{-\Phi(z)} F^2,  \label{wirkung}
  \end{equation}
where $F^2 = g^{MM'}g^{NN'} F_{MN}F_{M'N'}$ with $F_{MN} = \partial_M V_N - \partial_N V_M$ ($M,N = 0, \ldots, 4$) denoting the field strength tensor of the $U(1)$ vector field $V$, where the gauges $V_4=0$ and $\partial^\mu V_\mu = 0$ can be applied by Klein-Kaluza decomposition. $V_{\mu}$ is then sourced by a current operator $\bar q \gamma_{\mu} q$ in the boundary theory in the spirit of the field-operator correspondence. $z$ is the holographic bulk coordinate; the determinant of the metric $g_{MN}$ is $g$, and $k=12\pi^2L/N_c$ stands for the gauge coupling with the AdS radius $L$. Greek indices run in the range $0 \ldots 3$. The vector field $V_{\mu}= \epsilon_\mu \varphi(z) \exp( i p_{\nu}  x^{\nu} )$ has the polarization described by $\epsilon_{\mu}$; the important part of the wave function is the amplitude $\varphi(z)$ beyond the phase $p_{\nu}x^{\nu}$. The metric, now asymptotically AdS with a black hole embedded optionally, can be read off the infinitesimal line element squared
  \begin{equation}
   \dd s^2 = e^A \li(f(z) \dd t^2 -\dd \vec x^{\,2} - \frac{\dd z^2}{f(z)}\ri). \label{ds}
  \end{equation}
The resulting equation of motion can be cast into the form of a one-dimensional Schr\"odinger equation
  \begin{equation}
   \li(\Dp_{\xi}^2 -(U_T-m_n^2) \ri) \psi=0 \label{schroe}
  \end{equation}
by a coordinate transformation $z \to \xi$ with $\dd \xi (z) / \dd z= 1 /f(z)$ and the transformation $\psi =  \varphi \, \exp\{-\frac{1}{2}(A - \Phi)\} $. Note that $k$ drops. The potential reads
  \begin{equation}
   U_T = \li(\frac12 (\frac12 \Dp_{z}^2 A-\Dp_{z}^2 \Phi) +\frac14 (\frac12 \Dp_{z} A-\Dp_{z} \Phi)^2 \ri) f^2+ \frac14 (\frac12 \Dp_{z} A-\Dp_{z} \Phi) \Dp_{z} f^2. \label{hotpot}
  \end{equation}
The case $T=0$ is equivalent to setting $f=1$. \\
The vector meson states correspond to normalizable solutions of (\ref{schroe}) with masses squared $m_n^2=p_{\mu}p^{\mu}$. Since the field $V_{\mu}$ is considered in the probe limit, i.e.~the remaining warp factor $A(z)$ and the dilaton profile $\Phi(z)$ are adjusted to recover at $T=0$ a Regge type spectrum $m_n^2= \beta_0+\beta_1n+\beta_2n^2 + \ldots$ with $\abs{\beta_{n>1}} \ll \beta_{0,1}$ and $n=0,1,2, \ldots$ (see \cite{KKSS, Brodsky, KZ, Bugg} for discussions of the Regge trajectories). In line with the original SW model \cite{KKSS}, an easy ansatz is $A(z)=-2\ln (z/L)$ and $\Phi(z)=(cz)^p$. We put $L=1/c$ and obtain $c=291$ MeV, $p=3.77$ for the $\rho$ meson and the first three radial excitations, see Fig.~\ref{p3_abb1}-left. The ground state (g.s.) and the first two excitations are in fact on a linear trajectory with $\beta_0=0.49$ GeV$^2$ and $\beta_1=1.34$ GeV$^2$; the third excited state suffers from some modest offset. Such a spectrum is advocated in \cite{Ebert}. Reference \cite{BK}, in contrast, employs another sequence of excitations with $\beta_0=0.68$ GeV$^2$ and $\beta_1=0.72$ GeV$^2$ which can be captured very well up to $n=6$ by $c=399$ MeV and $p=2.09$, see Fig.~\ref{p3_abb1}-right. These latter values are fairly well in agreement with the scale setting and quadratic dilaton profile in the original SW model \cite{KKSS}. \\
In such a way, the hadron energy scale $c$ is fixed by properties of the considered hadron spectrum. This highlights also the role of the dilaton field as conformal symmetry breaker.

\section{Thermodynamics and setting of $\boldsymbol{\tc}$}
\begin{sidewaysfigure}
   \cen{\includegraphics[scale=.62]{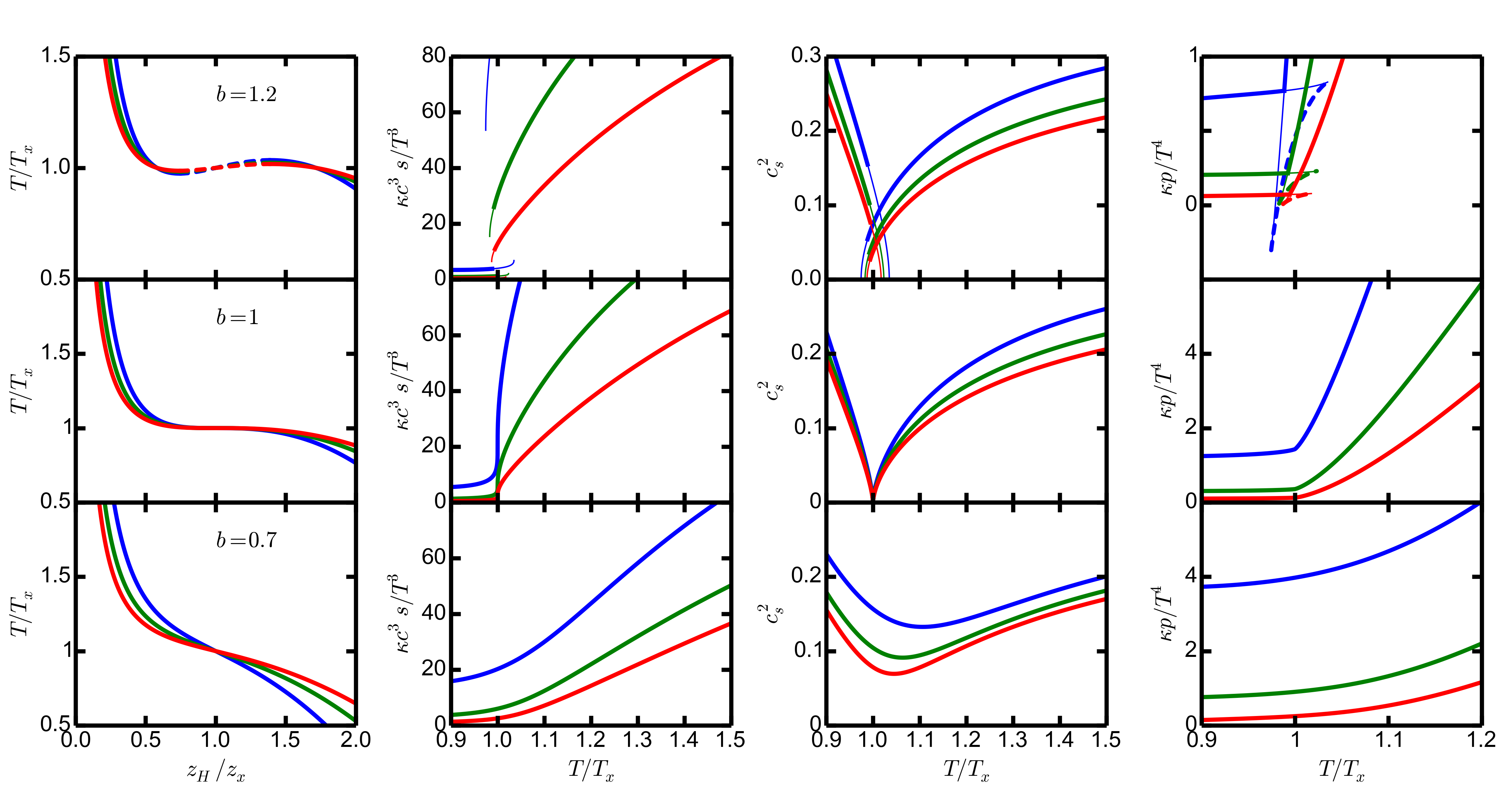} % oder 0.1475
   \caption{Characterization of the thermodynamics determined by Eq.~(\ref{T}). Left column: $T/T_x$ vs.~$z_H/z_x$, next column: scaled entropy densities as a function of $T/T_x$, next but one column: velocities of sound squared as a function of $T/T_x$, right column: scaled pressures as a function of $T/T_x$. Top row: $b=1.2$, middle row: $b=1$, bottom row: $b=0.7$. Solid curves: stable branches, thin curves: metastable branches, dashed curves: unstable branches; red/blue/green curves: $\theta=\frac23 \pi / \pi / \frac43 \pi$.} \label{p3_abb2}}
  \end{sidewaysfigure}

It is important to note that in the extension to non-zero temperatures we retain the above quoted warp factor $A(z)$. This implies, in the presence of a black brane, the Bekenstein-Hawking entropy density is $s(z_H)=(2\pi / \kappa) \exp \{\frac32 A(z_H)\}$ (the constant $\kappa$ is related to the five-dimensional gravitational constant $G_5$ via $\kappa=8\pi G_5$), meaning $s \propto 1/z_H^3$. To have asymptotically an AdS geometry, one requires $f(z\to 0)=1$ and $\Dp^k_zf(z) \mid_{z\to0}=0$ for $k=1,2,3$, i.e.~$z \to 0$ is the AdS boundary in the coordinate system (\ref{ds}). The black brane is facilitated by a simple zero $f(z=z_H)=0$, implying $T(z_H)= -4\pi \Dp_z f(z) \mid_{z=z_H}$. Both, the Bekenstein-Hawking entropy density $s(z_H)$ and the Hawking temperature $T(z_H)$ combine to the equation of state in parametric form, resulting finally in $s(T)$. At vanishing net baryon density, general thermodynamics dictates for the pressure 
$p(T) = \int \nolimits_{T_0}^T \dd \vartheta s (\vartheta) +p_0$, where $p_0$ is a scale set at temperature $T_0$. The velocity of sound squared is $c_s^2= \Dp \ln T / \Dp \ln s$; it has the advantage to be free of any further scale setting for $L$ and $c$. \\
Our ansatz for the blackness function is
  \begin{equation}
   f(z) = 1-\frac{z^4}{z_H^4} \li(1+ \frac{2(\pi z_H T(z_H)-1)}{\exp\{\frac2e (\pi z_H T(z_H)-1)\}} \li[ \li(\frac{z}{z_H} \ri)^{2 \exp\{\frac2e (\pi z_H T(z_H)-1)\}}-1\ri] \ri) \label{f}
  \end{equation}
(see appendix in \cite{ich2}). It is, of course, not unique, but constructed in such a manner to meet the above criteria for $f(z)$. What remains is a suitable model for $T(z_H)$. In agreement with the above quoted three QCD related options, as offered by the Columbia plot in \cite{Philipsen:2015eya}, we utilize:
  \begin{equation}
  \frac{T(z_H)}{T_x} =\frac1{\theta x} + 1 - \frac{3b}{\theta}+ \frac{3bx}{\theta} - \frac{x^2}{\theta}, \label{T}
  \end{equation}
where $\theta=\pi z_xT_x$, $x=z_H/z_x$, and $T_x$ sets the thermodynamic scale and $\theta$ (or $z_x$) is a parameter. As shown below, the dimensionless parameter $b$ steers the order of the transition from the high-temperature (plasma or deconfined) phase to the low-temperature (hadron or confined) phase.\\
Graphs of $T/T_x$ vs. $z_H/z_x$ (left column), $\kappa c^3 s/T^3$ (second column), $c_s^2$ (third column) and $\kappa p/T^4$ (right column) are exhibited in Fig.~\ref{p3_abb2} for three values of the parameter $\theta=\frac23$ (blue curves), 1 (green curves) and $\frac43$ (red curves). We have chosen several ad hoc values of $p_0/T_x^4$ to shift the curves of the scaled pressure (right column). For $b>1$, the curves $T/T_x$ vs. $z_H/z_x$ (left top panel) display a local minimum and a local maximum; the scaled pressure obeys the loop with a self intersection at $\tc$, as characteristic for a first-order phase transition. The unstable sections are shown by dashed curves, and the thin curves are for the metastable branches, also used for the scaled entropy density and squared sound velocity (middle panels). The solid curves are for the stable branches. In contrast to the Hawking-Page transition and the related thermal gas solution (see Appendix), below $\tc$, the medium has a non-zero sound velocity, entropy density and pressure as exhibited in the top row. One can tentatively attribute the medium at temperatures above $\tc$ to a plasma (deconfined) state, while below $\tc$ it would correspond to a hadron (confined) gas state, as stressed above. Since the present holographic approach does not have some analog of the running QCD coupling, we refrain to study the high-temperature region. Analogously, too far below $\tc$ the rich hadron physics is probably also not accessible in such a simple model. That is why we confine ourselves to the region around $\tc$.\\
For $b=1$ (middle row), the temperature $T/T_x$ vs.~$z_H/z_x$ exhibits a stationary turning point at $z_H=z_x$; the pressure loop just vanishes, and the sound velocity vanishes at $\tc=T_x$. These features are characteristic for a second-order phase transition, which is at the demarcation of first-order phase transition and cross over according to the Columbia plot. \\
For $b<1$, the temperature drops continuously with increasing $z_H$ (left bottom panel), and scaled entropy density and pressure are monotonously increasing with temperature (see bottom row). Still, the velocity of sound exhibits minima, corresponding to softest points of the medium. We use these minima to define a $\tc$, with the meaning of `cross over'. The ordering of the various curves for different values of $\theta$ in Fig.~\ref{p3_abb2} does not change with $b$.

\begin{figure}
   \cen{\includegraphics[scale=.7]{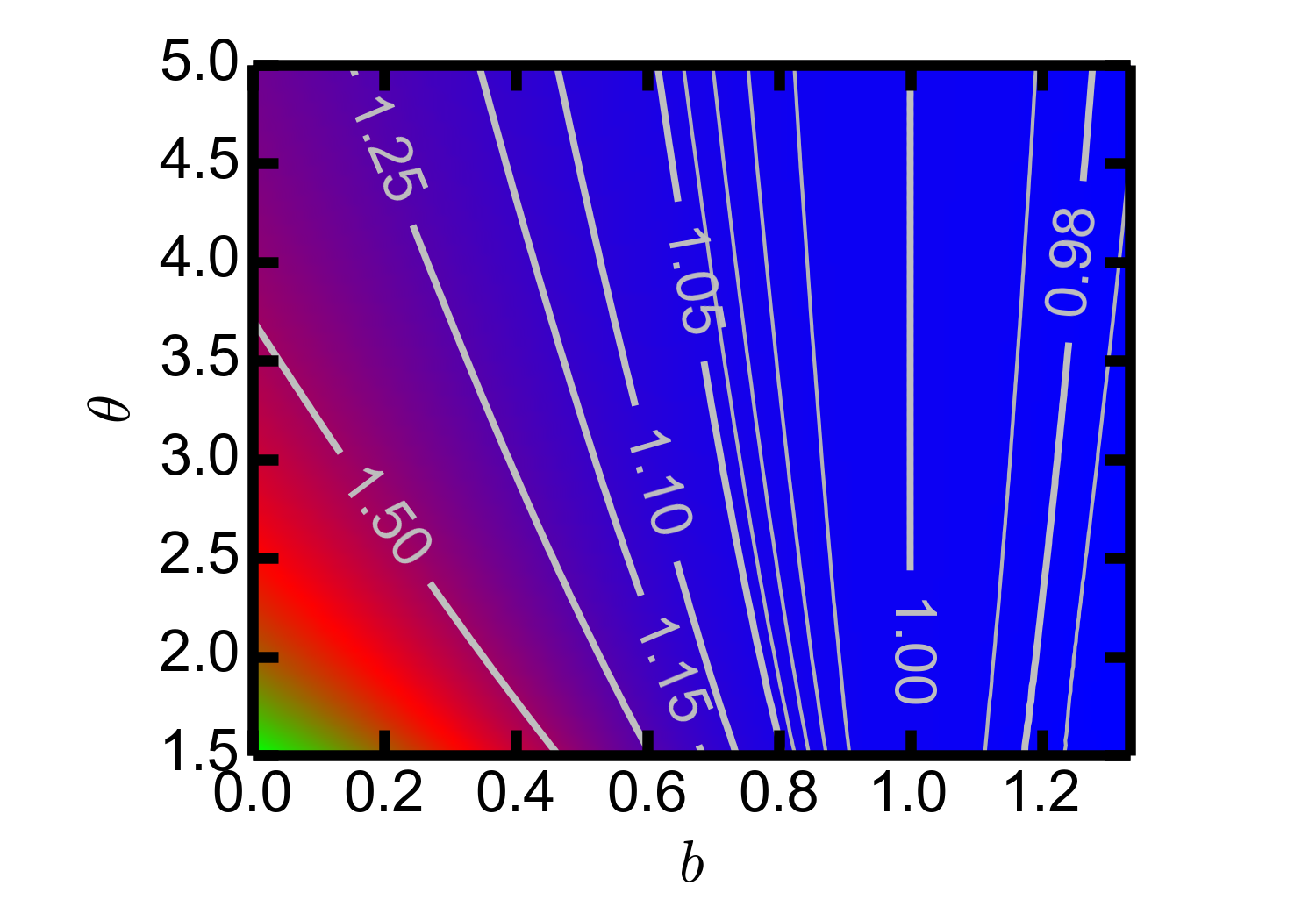}
   \caption{Contour plot of $\tc/T_x$ over the $\theta$-$b$ plane. $\tc$ is either determined by the first-order phase transition ($b>1$) or the second-order phase transition (minimum of sound velocity, $b=1$) or the minimum of the sound velocity ($b<1$).} \label{p3_abb3}}
  \end{figure}

Figure \ref{p3_abb3} exhibits a contour plot of the above defined values of $\tc$ scaled by $T_x$ above the $\theta$-$b$ plane. Thinking of a scale setting of $T_x=150$ MeV, this figure unravels that $\tc$ is about $T_x$ for $b$ uncovering the interval $0.5 \ldots 1.5$, thus accommodating the QCD relevant cases of first-order ($b>1$), second-order ($b=1$) transitions as well as the cross over ($b<1$) for not too large values of $\theta$.

\section{Appearance of vector mesons at $\boldsymbol{T_{\rm dis}}$}
\begin{figure}
   \cen{\includegraphics[scale=.48]{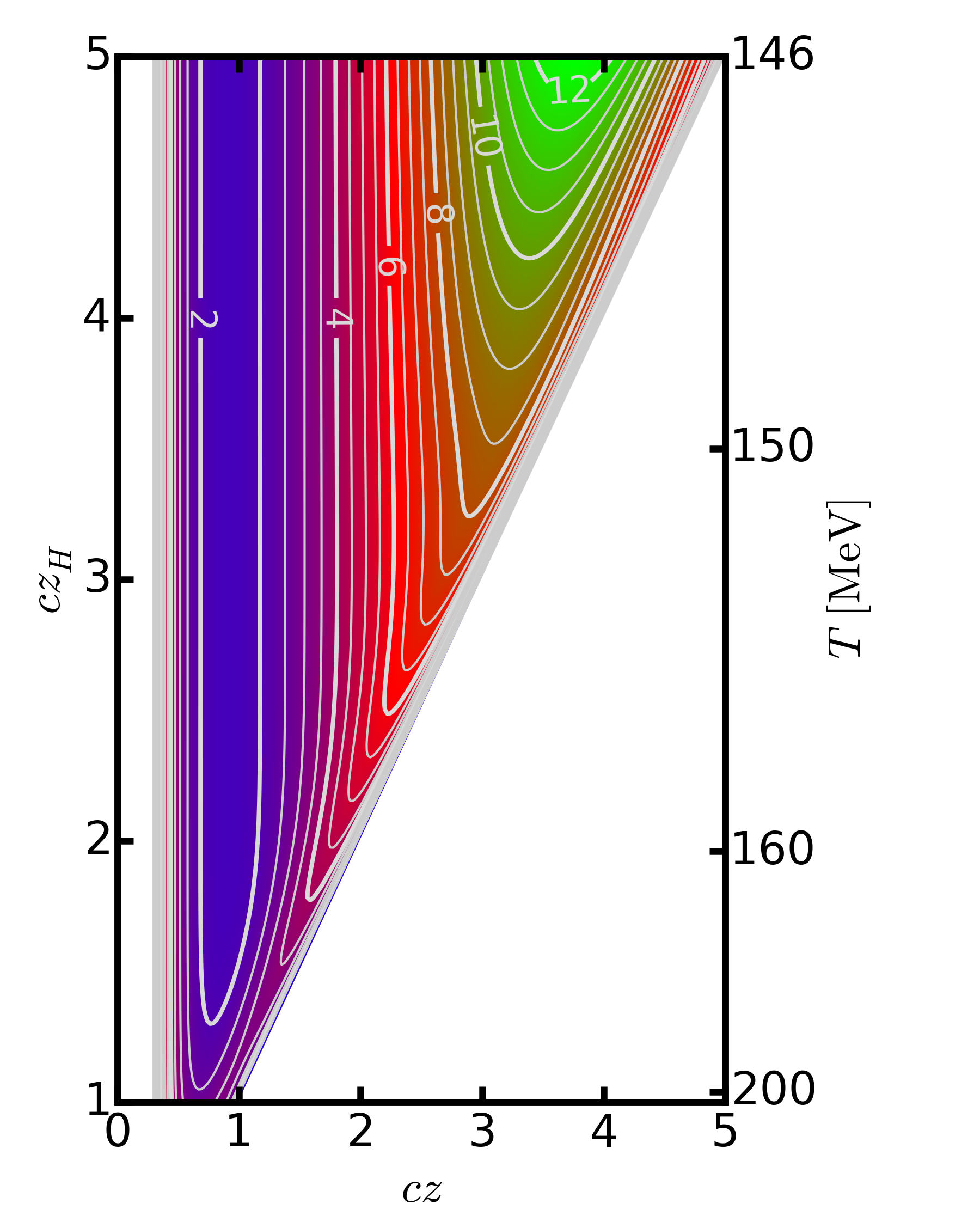}  \includegraphics[scale=.113]{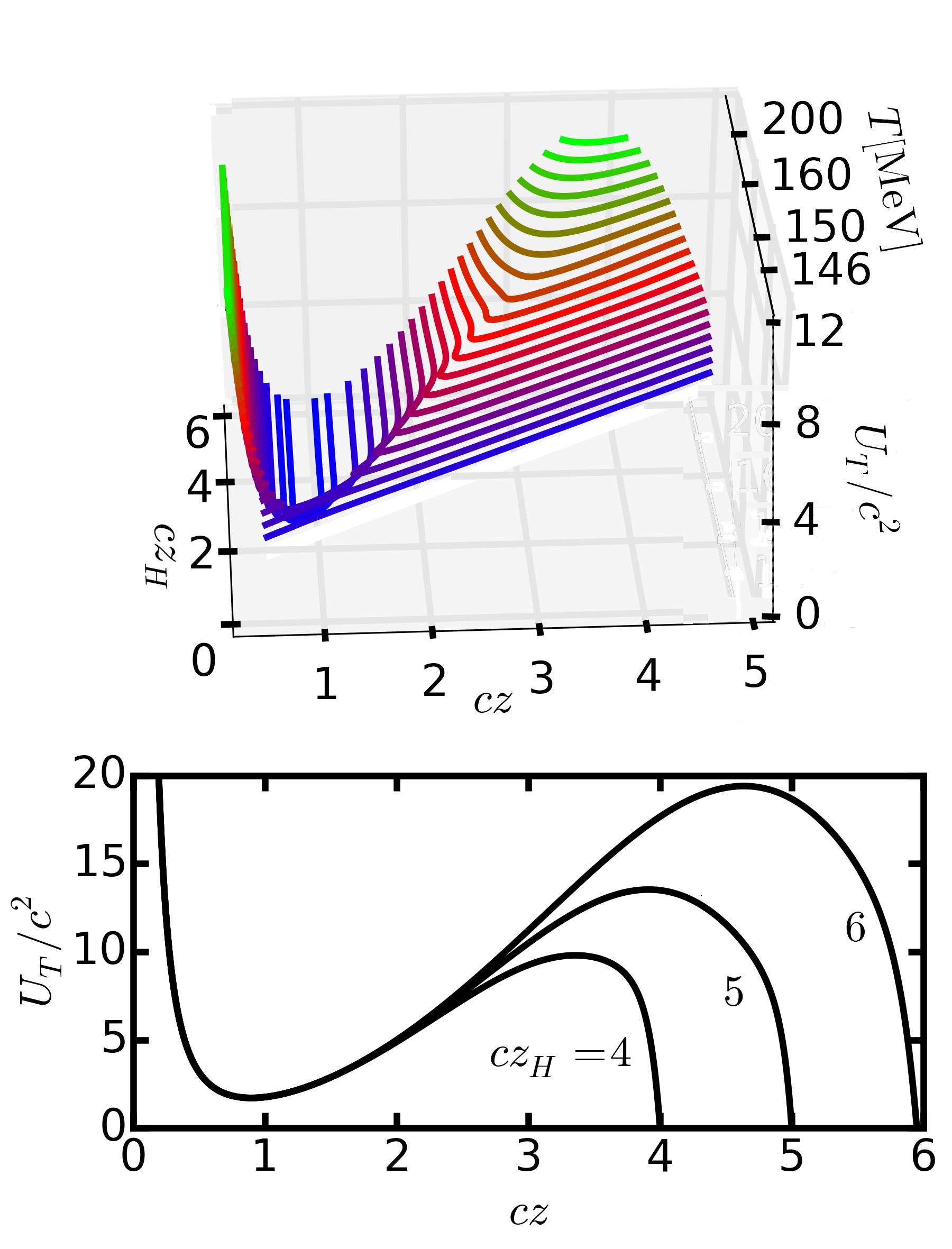}
   \caption{An example of the potential $U_T(z,z_H)/c^2$ over the $z$-$z_H$ plane for the parameters $p=2.09$, $c=399$ MeV, $T_x=150$ MeV, $\theta=2.3$ and $b=0.94$. Left panel: contour plot, right panels: 3D view with contours $U_T={\rm constant}$ (top) and cross sections $U_T(z,z_H={\rm constant})$ for various values of $z_H$ (bottom). The l.h.s~infinite wall at $z \to 0$ is hardly influenced by temperature effects.} \label{p3_abb4}}
  \end{figure}

After setting the thermodynamic scale by fixing $T_x$ we turn to the impact of non-zero temperatures on the spectrum of vector mesons. In contrast to thermodynamics, where only $T(z_H)$ matters (together with $s \propto 1/z_H^3$), now the explicit form of the blackness function Eq.~(\ref{f}) becomes important. To get an impression of the shape of the potential Eq.~(\ref{hotpot}) we exhibit in Fig.~\ref{p3_abb4} view graphs of $U_T(z,z_H)$. Due to temperature effects the asymmetrically U-shaped potential $U_{T=0}$ becomes at the r.h.s.~strongly deformed. It is the zero of $f$ at the horizon $z=z_H$ which bends also $U_T$ to zero there. The wall nearby $z<z_H$ becomes quadratically lower with smaller (larger) values of $z_H$ ($T$), implying that less states as normalizable solutions of Eq.~(\ref{schroe}) can be accommodated. In addition, the states still existing get a width due to a finite barrier penetration probability. At a certain value of $z_H$ (or $T$), the r.h.s.~wall is so low that even the ground state cannot be longer accommodated. This is our definition of $T_{\rm dis}^{\rm g.s.}$. At $T>T _{\rm dis}^{\rm g.s.}$ hadron states do not exist at all. As launched in \cite{ich1} we consider this as an emulation of deconfinement, here proven only for the vector channel. It is now a matter of the steepness of the ramp of the r.h.s.~wall of $U_T$ whether all hadron states (dis)appear at once or gradually, i.e. $T_{\rm dis}^{\rm g.s.}=T_{\rm dis}^{\rm 1^{st}} = \ldots$ or $T_{\rm dis}^{\rm g.s.}> T_{\rm dis}^{\rm 1^{st}} > \ldots$, where $\rm 1^{st}$ labels the first excited state. 

\begin{figure}
   \cen{\includegraphics[scale=.79]{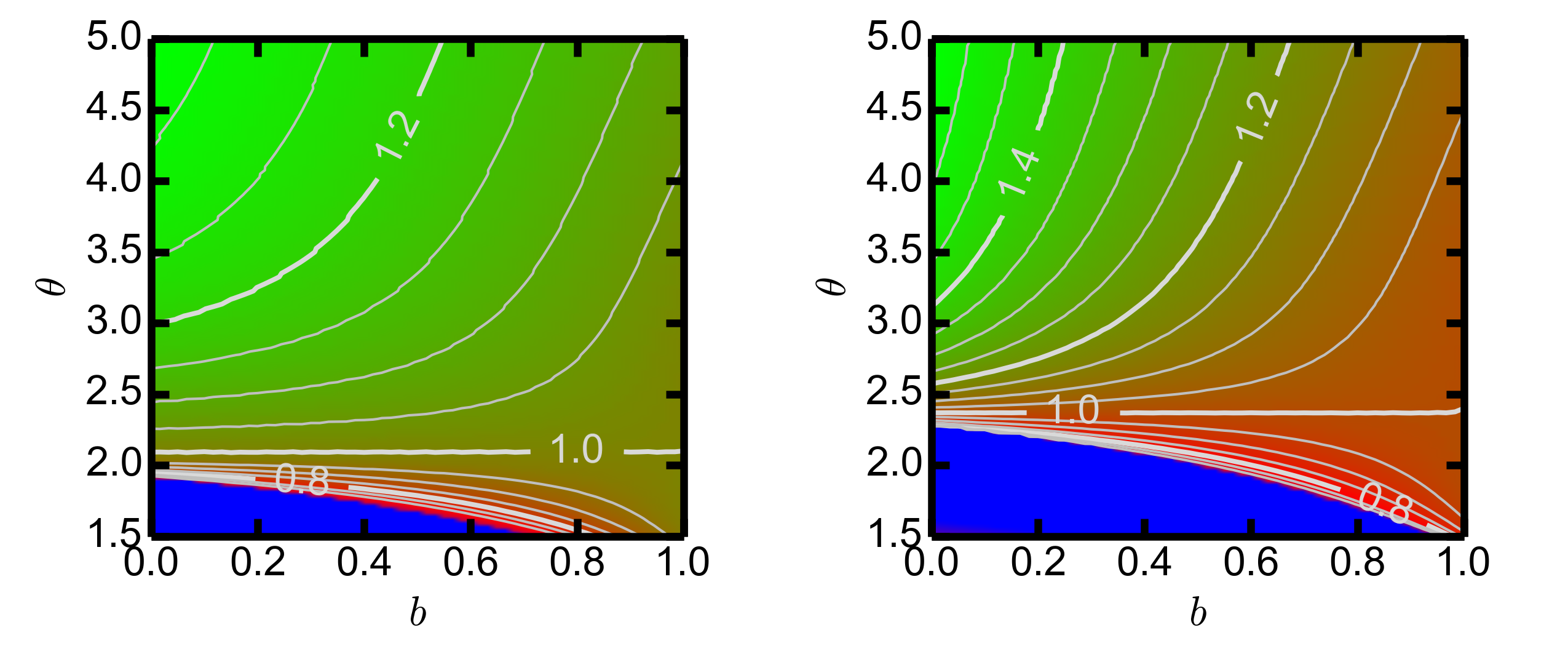} 
   \caption{Contour plot of $T_{\rm dis}^{\rm g.s.}/T_x$ over the $\theta$-$b$ plane for $T_x=150$ MeV and for parameter choices as in Fig.~\ref{p3_abb1}.} \label{p3_abb5}}
  \end{figure}

Figure \ref{p3_abb5} quantifies $T_{\rm dis}^{\rm g.s.}/T_x$ as a contour plot over the $\theta$-$b$ plane for $T_x=150$ MeV, as identified in the above scale setting for QCD relevant thermodynamics. One infers from that figure that a suitable value of $\theta$ can be chosen to arrive at $T_{\rm dis}^{\rm g.s.}=T_x$. We focus here on the region $b \leq 1$, as most relevant for mimicking QCD features. Variations of $T_x$ by $\pm (10 \ldots 20)$ MeV do not change the pattern displayed in Fig.~\ref{p3_abb5}. 

\begin{figure}
   \cen{\includegraphics[scale=.79]{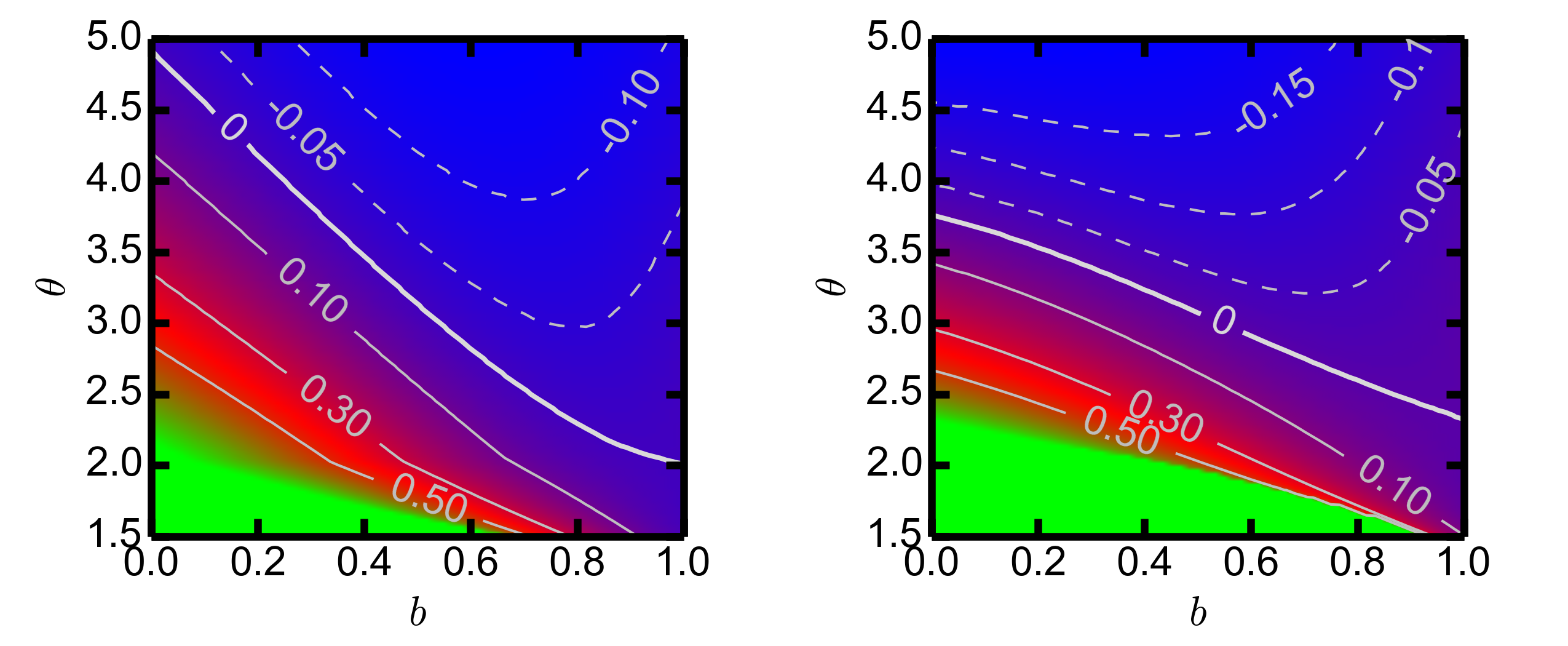} 
   \caption{Contour plot of $(\tc-T_{\rm dis}^{\rm g.s.})/T_x$ over the $\theta$-$b$ plane for $T_x=150$ MeV and for parameter choices as in Fig.~\ref{p3_abb1}.} \label{p3_abb6}}
  \end{figure}

The final step is to arrive at $T_{\rm dis}^{\rm g.s.}=\tc=150$ MeV. For quantifying the options we exhibit in Fig.~\ref{p3_abb6} the quantity $(\tc-T_{\rm dis}^{\rm g.s.})/T_x$ as contour plot over the $\theta$-$b$-plane, again focusing on $b\leq 1$. The fat gray curve with label 0 is in fact the $\tc-T_{\rm dis}^{\rm g.s.}=0$ locus. That is, one can indeed synchronize the thermodynamics at $\tc$ with the deconfinement emulation. 

\begin{figure}
   \cen{\includegraphics[scale=.7]{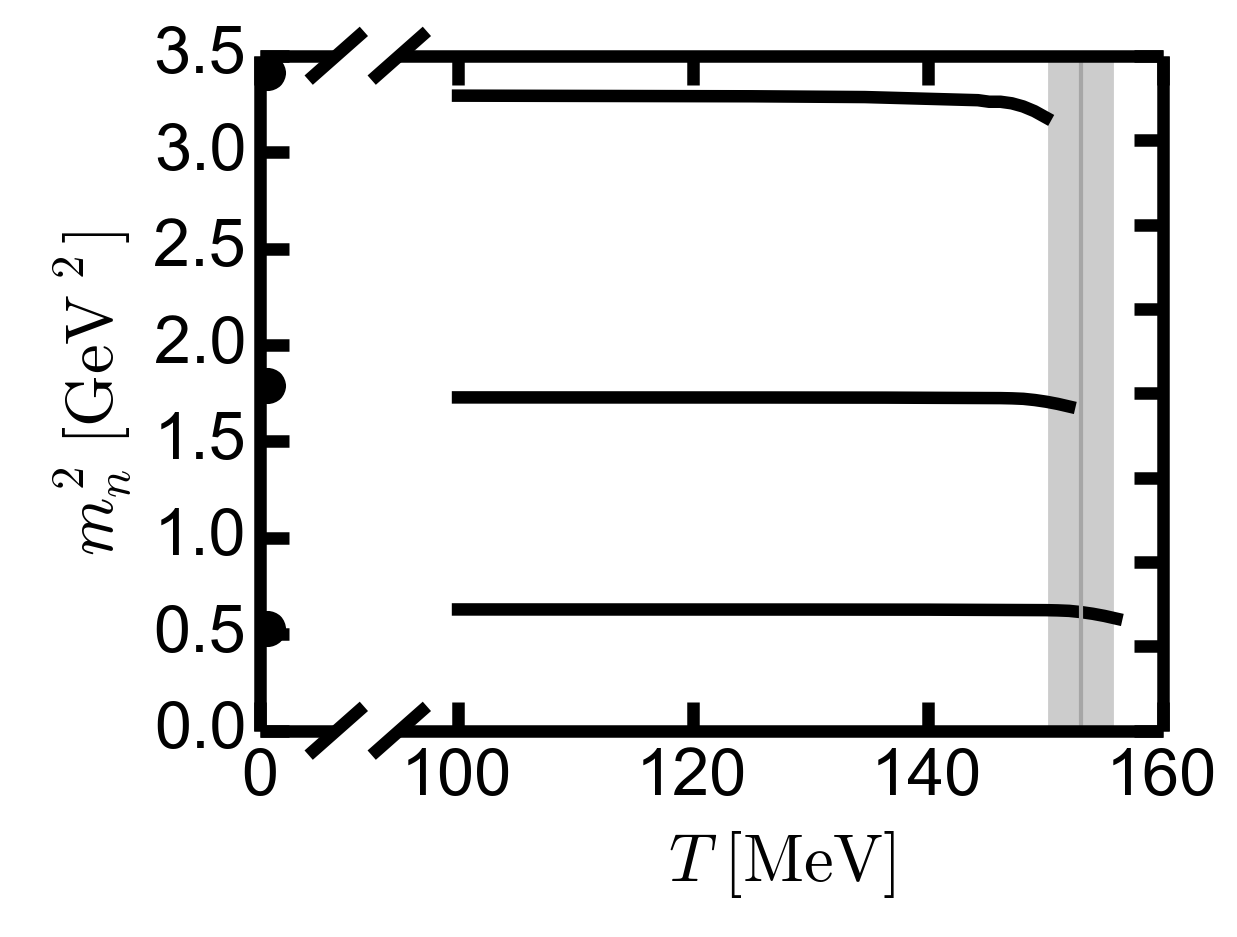} \includegraphics[scale=.7]{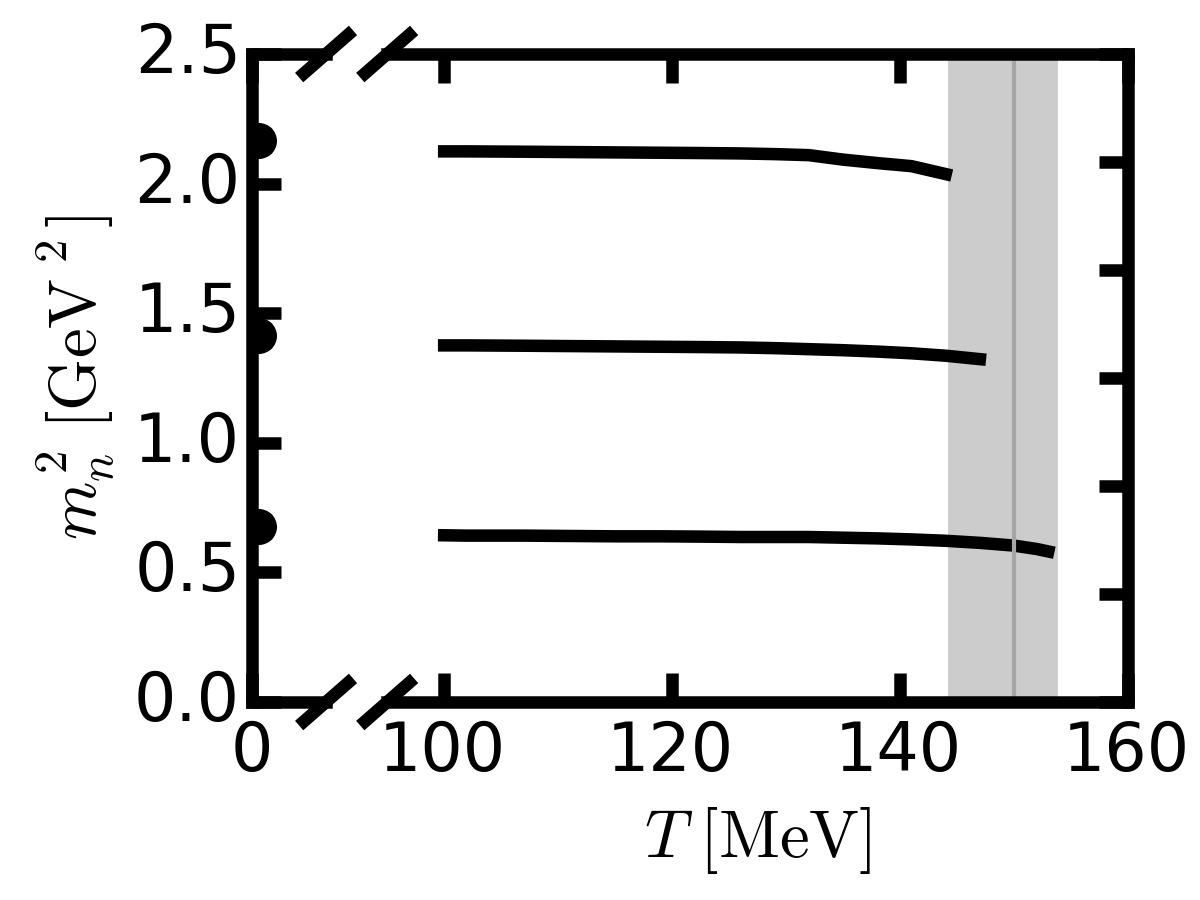}
   \caption{The first three vector meson masses squared as a function of the temperature for parameters $p=3.77$ and $c=291$ MeV (left panel) or $p=2.09$ and $c=399$ MeV (right panel) as in Fig.~\ref{p3_abb1}. The additional parameters for both panels are $T_x=150$ MeV, $\theta=2.3$ and $b=0.94$. For these parameters corresponding to the QCD relevant cross over, sequential hadronization happens in the vertical gray corridor centered at $\tc$ (vertical line). Bullets depict the values of $m_n^2$ at $T=0$.} \label{p3_abb7}}
  \end{figure}

Figure \ref{p3_abb7} exhibits the temperature dependence of the lowest three vector meson states. There is a very weak temperature dependence, and the states appear in a narrow corridor centered at $\tc$ upon cooling. The width of the hadronization corridor can be dialed by selecting other parameters $\theta$ and $b$, e.g.~for $b=1$ and $\theta=2.5$ it shrinks to 3 MeV. In contrast to the finding in \cite{ich1}, here the impact of increasing temperature is a dropping of the masses near $\tc$. We can trace back that difference to details of the function $f(z)$ near $z_H$: Even for the same slope of $f(z=z_H)$, the slopes of $f(z=z_H[1-\ve])$ can differ ($\ve$ is a small number.). The net effect is that the r.h.s.~of $U_T$ is squeezed or stretched relative to $U_{T=0}$ with the implication of kicking up or down $m_n(T\approx \tc)$. For $z\ll z_H$, $U_T$ suffers only from minor modifications relative to $U_{T=0}$, see Fig.~\ref{p3_abb4}. That is the reason for $m_n(T<\tc) \approx {\rm constant}$ as seen in Fig.~\ref{p3_abb7}.

Figure \ref{p3_abb7} unravels also the option of a sequential hadronization: $T_{\rm dis}^{\rm g.s.} > T_{\rm dis}^{\rm 1^{st}}>T_{\rm dis}^{\rm 2^{nd}}$ would imply $T_{\rm f.o.}^{\rm g.s.} > T_{\rm f.o.}^{\rm 1^{st}}>T_{\rm f.o.}^{\rm 2^{nd}}$ in a `strict freeze-out model' wherein hadron species are statistically distributed just at the maximum temperature at which they exist. Interestingly, such a refinement of the thermo-statistical freeze-out models is currently matter of debate. For instance, \cite{Bellwred} disputes a data-enabled scenario with lower freeze-out temperature of light-quark dominated hadrons in comparison with heavy-flavor dominated hadrons. Our present approach, however, is restricted to the light-quark sector and must not be related to the multi-flavor scenario in \cite{Bellwred}. In addition, considering hadron states above the ground state as resonances, a plethora of potentially obscuring effects needs to be considered, such as regeneration reactions in the confined phase and issues of the (experimental) reconstruction of rapidly decaying resonances in the hadronic environment which distorts the decay products and obstructs the reconstruction of the parent resonances \cite{Ilner, Knospe}. These facets call for an improvement of understanding the hadronization dynamics during the QCD cross over transition in relativistic heavy-ion collisions. For further discussion and citations of hadronization issues in the particularly interesting charm sector we refer the inquisitive reader to \cite{Petri}.

\section{Summary}
In summary we employ an extended soft wall model with the following properties:
(i) approximately linear Regge trajectory of the first few excitations of the $\rho$ meson,
(ii) various options of the thermodynamics related to QCD (e.g.~cross over or second-order phase transition or first-order phase transition),
(iii) disappearance of vector meson states for temperatures larger than the cross over or transition temperatures.
In the particularly important case of the cross over, the relevant temperature scale is about $\tc \approx 150$ MeV, both for thermodynamics and the emulation of deconfinement as disappearance of hadron states. Such a holographic model - even quite schematic and restricted to vector mesons in the light quark sector - is consistent with the thermo-statistical model analyses of hadron yields since the hadron states at $T<\tc$ are as the spectrum at $T=0$, with the exception of tiny temperature effects in a narrow corridor centered at $\tc$, meaning no noticeable medium effects below $\tc$. \\
Obvious refinements within the given framework should address other Regge trajectories (e.g.~heavy-quark vector mesons \cite{Braga, Col11, Park} and heavy-light quark vector mesons) up to pseudo-scalar, scalar, axial-vector and tensor channels, thus generalizing the model to many more degrees of freedom (both beyond vector mesons and the dilaton `dynamics') towards the goal to source with them in a dynamical way the holographic gravity beyond the probe limit and include back reactions. For extensions to non-zero net baryon density, the baryons and related order parameters must be included too. 

\begin{acknowledgments}
The work is supported by BMBF and Studienstiftung des deutschen Volkes. 
\end{acknowledgments}  

\begin{appendix}
\section{Hawking-Page transition}
\begin{sidewaysfigure}
   \cen{\includegraphics[scale=.61]{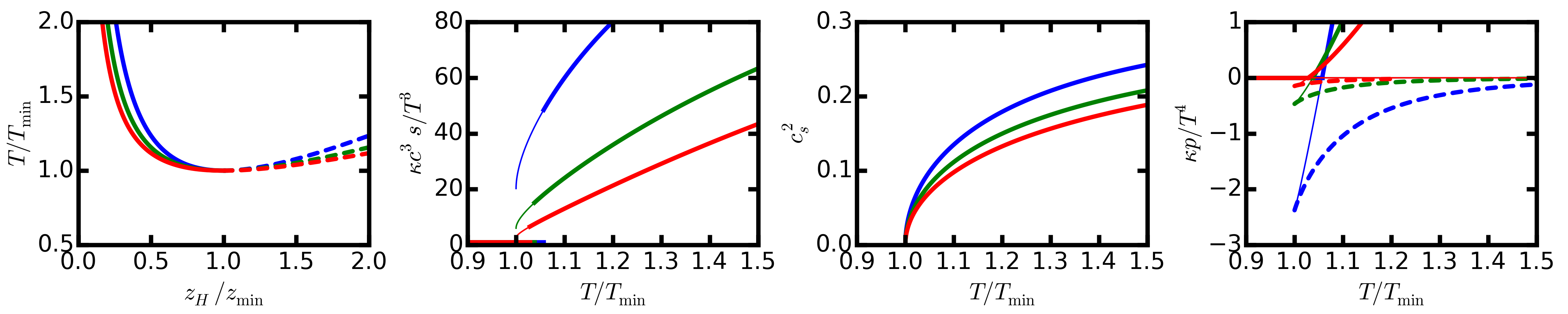}
   \caption{As Fig.~\ref{p3_abb2}, however for the temperature vs.~horizon-position relation (\ref{a1}).} \label{p3_abbA}}
  \end{sidewaysfigure}

We supplement here the thermodynamics for the model considered in \cite{ich1} for 
\begin{equation}
   \frac{T(z_H)}{T_{\min}}= \frac1{\hat \theta \hat x} + 1 -\frac2{\hat \theta}+ \frac{\hat x}{\hat \theta} \label{a1} ,
  \end{equation}
where, instead of $T_x$, $z_x$, we now consider $T_{\min}$ and $z_{\min}$ as parameters, and use here $\hat x=z_H/z_{\min}$, $\hat \theta=\pi z_{\min}T_{\min}$. $T(z_H)$ displays a global minimum at $T_{\min}$ and $z_{\min}$. \\
In the spirit of Fig.~\ref{p3_abb2}, the scaled temperature, velocity of sound and scaled pressure are exhibited in Fig.~\ref{p3_abbA}. The pressure loop construction is special here: $p(T(z_H \to \un)) =0$ is an element thereof, i.e.~$p_0=0$ and $T_0=\un$. In the thermal gas (low-temperature) phase the pressure is assumed to scale with $N_c^0$, where $N_c$ is number of colors, while in the black hole (high-temperature or plasma) phase it goes as $N_c^2$, i.e. $p \approx 0$ represents the thermal gas \cite{Kiritsis}. Accordingly, below $\tc$ - determined by $p=0$ of the black hole solution - there is a nearly pressure-free and nearly zero-entropy `medium', where also the sound velocity is near zero. Obviously, such a construction relying on the Hawking-Page transition is less adequate for describing qualitatively QCD features for realistic quark masses.

\end{appendix}


\begin{thebibliography}{99}
\bibitem{KKSS} A. Karch, E. Katz, D. T. Son, M. A. Stephanov, Phys. Rev. D 74, 015005 (2006).
\bibitem{Maldacena} J. M. Maldacena, Adv. Theor. Math. Phys. 2, 231 (1998).
\bibitem{Witten} E. Witten, Adv. Theor. Math. Phys. 2, 505 (1998).
\bibitem{Gubser} S. S. Gubser, I. R. Klebanov, A. M. Polyakov, Phys. Lett. B 428, 105 (1998).
\bibitem{Col12} P.  Colangelo, F. Giannuzzi, S. Nicotri, JHEP 1205, 076 (2012).
\bibitem{Col09} P. Colangelo, F. Giannuzzi, S. Nicotri, Phys. Rev. D 80, 094019 (2009). 
\bibitem{Gherghetta} T. Gherghetta, J. Kapusta, T. Kelley, Phys. Rev. D 79, 076003 (2009).
\bibitem{BK} S. P. Bartz, J. I. Kapusta, Phys. Rev. D 90, 074034 (2014).
\bibitem{Rapp1} R. Rapp, H. van Hees, Eur. Phys. J. A 52, 257 (2016).
\bibitem{Rapp2} R. Rapp, H. van Hees, Phys. Lett. B 753, 586 (2016).
\bibitem{Leu} S. Leupold, V. Metag, U. Mosel, Int. J. Mod. Phys. E 19, 147 (2010).
\bibitem{Hay} R. Hayono, T. Hatsuda, Rev. Mod. Phys. 82, 2949 (2010).
\bibitem{Fri} B. Friman et al. (Eds.), Lecture Notes in Phys. 814, 1 (2011).
\bibitem{Fuk} K. Fukushima, C. Sasaki, Prog. Part. Nucl. Phys. 72, 99 (2013).
\bibitem{Hilger} T. Hilger, R. Thomas, B. K\"ampfer, S. Leupold, Phys. Lett. B 709, 200 (2012).
\bibitem{Hohler} P. Hohler, R. Rapp, Phys. Lett. B 731, 103 (2014).
\bibitem{Bartz-Jacobson} S. P. Bartz, T. Jacobson, Phys. Rev. D 94, 075022 (2016).
\bibitem{ich1} R. Z\"ollner, B. K\"ampfer, Phys. Rev. C 94, 045205 (20016).
\bibitem{ich2} R. Z\"ollner, F. Wunderlich, B. K\"ampfer, arXiv:1611.04124 [hep-th] (2016).
\bibitem{Stachel2} A. Andronic, P. Braun-Munzinger, J. Stachel, Nucl. Phys. A 834, 237 (2010).
\bibitem{Stachel3} A. Andronic, P. Braun-Munzinger, K. Redlich, J. Stachel, arXiv:1611.01347 [nucl-th] (2016).
\bibitem{CKWX} J. Cleymans, B. K\"ampfer, M. Kaneta, S. Wheaton, N. Xu, Phys. Rev. C 71, 054901 (2005).
\bibitem{Bec2} F. Becattini et al., Phys. Rev. C 90, 054907 (2014).
\bibitem{thst1} G. Agakishiev  et al. (HADES Collaboration), Eur. Phys. J. A 52, 178 (2016).
\bibitem{thst3} L. Ferroni, F. Becattini, Acta Phys. Polon. B 43, 571 (2012).
\bibitem{Stachel} P. Braun-Munzinger, J. Stachel, C. Wetterich, Phys. Lett. B 596, 61 (2004).
\bibitem{Becattini} F. Becattini et al., Phys. Lett. B 764, 241 (2016).
\bibitem{Borsanyi2014} S. Bors$\rm \acute{a}$nyi et al., Phys. Lett. B 370, 99 (2014).
\bibitem{Bazavov2014} A. Bazavov et al., Phys. Lett. B 737, 210 (2014).
\bibitem{Philipsen:2015eya} C. Pinke, O. Philipsen, PoS LATTICE2015 (2016) 149.
\bibitem{Erdemenger} J. Erdmenger, M. Kaminski, F. Rust, Phys. Rev. D 77, 046005 (2008).
\bibitem{Erd2} J. Erdmenger, N. Evans, I. Kirsch, E. Threlfall, Eur. Phys. J. A 35, 81 (2008).
\bibitem{Erd3} J. Erdmenger, V. Fliev, JHEP 1101, 119 (2011).
\bibitem{Erd4} J. Erdmenger, N. Evans, M. Scott, Phys. Rev. D 91, 085004 (2015).
\bibitem{Cui11} M. Ammon, J. Erdmenger, Gauge/Gravity duality, Cambridge University Press (2015).   %L.-X. Cui, S. Takeuchi, Y.-L. Wu,  JHEP 1204, 144 (2012).
\bibitem{Brodsky} S. Brodsky et al., Phys. Rept. 584, 1 (2015).
\bibitem{KZ} E. Klempt, A. Zaitsev, Phys. Rept. 454, 1 (2007).
\bibitem{Bugg} D. Bugg, Phys. Rept. 397, 257 (2004).
\bibitem{Ebert} D. Ebert, R. N. Faustov, V. O. Galkin, Phys. Rev. D 79, 114029 (2009).
\bibitem{Bellwred} R. Bellwied, J. Phys. Conf. Ser. 736, 012018 (2016).
\bibitem{Ilner} A. Ilner et al., arXiv:1609.02778 [hep-ph] (2016).
\bibitem{Knospe} A. Knospe et al., Phys. Rev. C 93, 014911 (2016).
\bibitem{Petri} Y. Maezawa, P. Petreczky, Phys. Rev. D 94, 034507 (2016).
\bibitem{Braga} N. Braga, M. Contreras, S. Diles, Eur. Phys. J. C 76, 598 (2016).
\bibitem{Col11} P. Colangelo, F. Giannuzzi, S. Nicotri, Phys. Rev. D 83, 035015 (2011).
\bibitem{Park} C. Park, Phys. Rev. D 81, 045009 (2010).
\bibitem{Kiritsis} U. G\"ursoy, E. Kiritsis, L. Mazzanti, F. Nitti, JHEP 0905, 033 (2009).
\end{thebibliography}
\end{document}